\documentclass[12pt]{iopart}
\usepackage{amsfonts}
\usepackage{amssymb}
\usepackage{graphicx}
\usepackage{epsfig}
\usepackage{iopams}
\usepackage{hyperref}
\usepackage{cite}
\usepackage{color}

\begin{document}

\title{Quantum scars in spin-$1/2$ isotropic Heisenberg clusters}
\author{G. Zhang$^{1,2}$ and Z. Song{$^{\dagger}$}$^{1}$}

\ead{songtc@nankai.edu.cn}

\address{$^{1}$School of Physics, Nankai University, Tianjin 300071, China\\
$^{2}$College of Physics and Materials Science, Tianjin Normal University, Tianjin
300387, China}
\begin{abstract}
{We investigate the influence of the external fields on the statistics of
energy levels and towers of eigenstates in spin-$1/2$ isotropic Heisenberg
clusters, including chain, ladder, square and triangular lattices. In the
presence of uniform field in one direction, the SU($2$) symmetry of the
system allows that almost whole spectrum consists of a large number of
towers with identical level spacing. Exact diagonalization on finite
clusters shows that random transverse fields in other two directions drive
the level statistics from Poisson to Wigner-Dyson (WD) distributions with
different values of mean level spacing ratio, indicating the transition from
integrability to non-integrability. However, for the three types of
clusters, it is found that the largest tower still hold approximately even
the symmetry is broken, resulting to a quantum scar. Remarkably, the
non-thermalized states cover the Greenberger-Horn-Zeilinger (GHZ) and W
states, which maintain the feature of revival while a Neel state decays fast
in the dynamic processes. In addition, some dynamic schemes for experimental
detection are proposed. Our finding reveals the possibility of quantum
information processing that is immune to the thermalization in finite size
quantum spin clusters.}
\end{abstract}


\maketitle

\section{Introduction}

It is commonly believed that the main obstacle for the practical realization
of quantum information processing is the decoherence of the quantum state
caused by interactions with the environment. However, like a thermodynamic
system, where the process of thermalization always eventually destroys the
information of an initial state, the thermalization is also expected to be
unavoidable in a generic isolated nonintegrable\ quantum system. Recently,
it is well established that some nonintegrable systems can fail to
thermalize due to rare nonthermal eigenstates called quantum many-body scars
(QMBS) \cite%
{Shiraishi2017,Moudgalya2018,Moudgalya20182,Khemani2019,Ho2019,Shibata2020,McClarty2020,Richter2022,Jeyaretnam2021,Turner2018,Turner20182,Shiraishi2019,Lin2019,Choi2019,Khemani2020,Dooley2020,Dooley2021}%
. These nonthermal states are typically excited ones and span a subspace, in
which any initial states do not thermalize and can be back periodically. The
quantum many-body scarring can prevent the thermalization starting from
certain initial states. Therefore, the quantum information stored in the
subspace does not dissipate at finite temperature, holding promise for
potential applications in quantum information processing. The main task in
this field is finding scars in a variety of nonintegrable many-body systems.

In this work, we concentrate on quantum spin-$1/2$ Heisenberg clusters,
which have been successfully realized in experiments and studied under their
unitary time evolution\cite%
{Fukuhara2013,Fukuhara20132,Ronzheimer2013,Schneider2012,Cheneau2012,Jurcevic2014}%
. In this paper, our aim is to explore the transition from integrability to
non-integrability induced by external field and the possible quantum scars
in a simple Heisenberg model. In the presence of uniform field $h$ in $z$
direction, the SU($2$) symmetry of an isotropic Heisenberg system allows
that almost whole spectrum consists of large number of towers with identical
level spacing. It has been shown that the conjecture of Anderson
localization (AL)\cite{Anderson1958} can be extended to quantum spin systems
by applying random field, known as many-body localization (MBL)\cite%
{Znidaric2008,Pal2010,Vosk2013} which preventing thermalization and even
protecting quantum order\cite{Huse2013}. Here, we study the similar quantum
system, but focus on an alternative aspect.\ We investigate the influence of
the external fields on the statistics of energy levels and towers of
eigenstates in Heisenberg clusters, including chain, ladder, square and
triangular lattices. Exact diagonalization on finite clusters shows that
random transverse fields in $x$ and $y$ directions drive the level
statistics from Poisson to Wigner-Dyson (WD) distributions with two
different values of mean level spacing ratio. Numerical results show that
the cooperation between the uniform field $h$ in $z$ direction, and the
random field in $x$ or $y$ direction within their respective regions, takes
the crucial role for the transition from integrability to non-integrability.
Here we emphasize that the conclusion here is obtained only from small size
systems. But the number of energy levels is large enough to count its
statistics. For the three types of clusters, it is found that the largest
tower still hold approximately even the symmetry is broken, resulting to a
quantum scar. Remarkably, The non-thermalized states cover the
Greenberger-Horn-Zeilinger (GHZ) and W states, which maintain the feature of
revival while a Neel state decays fast in the dynamic processes. Our finding
reveals the possibility of quantum information processing that is immune to
the thermalization in finite size quantum spin clusters.

The remainder of this paper is organized as follows. In Sec. \ref{Model and
towers of eigenstates} we review the Heisenberg model and introduce the
towers of eigenstates. In Sec. \ref{Transitions of energy level statistics}
we perform the numerical computation of energy level statistics to
investigate the transition from integrability to non-integrability. In Sec. %
\ref{Quantum scars} we identify the quantum scars, surviving tower in the
presence of random field. We demonstrate the results by investigating the
dynamics of GHZ, W and Neel states in Sec. \ref{Revival of W and GHZ states}%
. Sec. \ref{conclusions} concludes this paper.

\section{Model and towers of eigenstates}

\label{Model and towers of eigenstates}

The system we study is a cluster of spin-$1/2$ isotropic Heisenberg model in
a random magnetic field with the Hamiltonian

\begin{equation}
H=H_{0}+H_{\mathrm{ran}},  \label{H}
\end{equation}%
which consists of two parts. The unperturbed system

\begin{equation}
H_{0}=\sum_{i,j\neq i}J_{ij}\mathbf{s}_{i}\cdot \mathbf{s}%
_{j}+h\sum_{j}s_{j}^{z},
\end{equation}%
and the perturbation term%
\begin{equation}
H_{\mathrm{ran}}=\sum_{j}\left( x_{j}s_{j}^{x}+y_{j}s_{j}^{y}\right) ,
\end{equation}%
where $s_{j}^{\lambda }$ ($\lambda =x,y,z$) are canonical spin-$1/2$
variables, and $\sum_{i,j\neq i}$ means the summation over all the possible
pair interactions at an arbitrary range. Here $\left\{ J_{ij}\right\} $ is
an arbitrary set of numbers, representing the strength of isotropic
spin-spin interaction. It only determines the structure of the system. For
simplicity, we set $J_{ij}=1$ or $0$ for the different cluster. It is
subjected to an external uniform field along the $z$-direction but random
fields along the $x$\ and $y$-direction. Here the field distribution is $%
x_{j}=$ran($-x,x$) and $y_{j}=$ran($-y,y$), where ran($-b,b$) denotes a
uniform random number within ($-b,b$).

We start with the case with $x=y=0$, which is the base of the rest study. We
review the construction of ferromagnetic states, and classify them into
different groups, referred as to towers of eigenstates\cite{Schecter2019}.
Due to the SU(2) symmetry of an isotropic Heisenberg model, we have%
\begin{equation}
\lbrack s^{\lambda
},H_{0}-h\sum_{j}s_{j}^{z}]=[s^{2},H_{0}-h\sum_{j}s_{j}^{z}]=0,  \label{SU2}
\end{equation}%
with the component of total spin operators%
\begin{equation}
s^{\lambda }=\sum_{j}s_{j}^{\lambda },
\end{equation}%
and%
\begin{equation}
s^{2}=\left( s^{x}\right) ^{2}+\left( s^{y}\right) ^{2}+\left( s^{z}\right)
^{2}.
\end{equation}%
The eigenstates of $H_{0}$\ can be expressed in the form $\left\vert \psi
_{n}(l,m)\right\rangle $, satisfying the eigen equations%
\begin{eqnarray}
H_{0}\left\vert \psi _{n}(l,m)\right\rangle &=&E_{n}(l,m)\left\vert \psi
_{n}(l,m)\right\rangle , \\
s^{2}\left\vert \psi _{n}(l,m)\right\rangle &=&l(l+1)\left\vert \psi
_{n}(l,m)\right\rangle , \\
s^{z}\left\vert \psi _{n}(l,m)\right\rangle &=&m\left\vert \psi
_{n}(l,m)\right\rangle ,
\end{eqnarray}%
where $l=N/2,N/2-1,...,0$\ and $m=l,l-1,...,-l$. Here $n$ is the index of
the towers, representing a group of eigenstates with equal energy level
spacing. Defining the tower operator%
\begin{equation}
Q^{\dag }=\sum_{i}\left( s_{i}^{x}+is_{i}^{y}\right) ,
\end{equation}%
we have
\begin{equation}
\lbrack H_{0},Q^{\dag }]=hQ^{\dag },[H_{0},Q]=-hQ
\end{equation}%
Then in each tower, the eigenstates have the relation%
\begin{eqnarray}
Q^{\dag }\left\vert \psi _{n}(l,m)\right\rangle &=&\sqrt{N}\left\vert \psi
_{n}(l,m+1)\right\rangle \\
Q\left\vert \psi _{n}(l,m)\right\rangle &=&\sqrt{N}\left\vert \psi
_{n}(l,m-1)\right\rangle
\end{eqnarray}%
and
\begin{equation}
E_{n}(l,m)\pm h=E_{n}(l,m\pm 1).
\end{equation}

Considering a cluster with even $N$ spins, the number of tower with $l=\frac{%
N}{2}-k$, with $k=0$, $1$, $2$,$...$,$\ N/2$, can be obtained as%
\begin{eqnarray}
N_{\mathrm{tower}}(0) &=&1, \\
N_{\mathrm{tower}}(k) &=&C_{N}^{k}-C_{N}^{k-1},k\neq 0
\end{eqnarray}%
Taking $N=12$ as an example, the structure of towers is illustrated in the
matrix consisting of diagonal blocks in Fig. (\ref{figure3}a1).\ In
addition, for an arbitrary cluster, the eigenstates of first tower can
always be expressed explicitly as%
\begin{equation}
\left\vert \psi _{1}(N/2,p-N/2)\right\rangle =\frac{1}{p!\sqrt{C_{N}^{p}}}%
\left( Q^{\dag }\right) ^{p}\left\vert \Downarrow \right\rangle ,
\end{equation}%
with energy%
\begin{equation}
E_{1}(N/2,p-N/2)=\frac{1}{4}\sum_{i,j\neq i}J_{ij}+\left( p-N/2\right) h,
\end{equation}%
for $p=0,...,N$, where $\left\vert \Downarrow \right\rangle =\left\vert \psi
_{1}(N/2,-N/2)\right\rangle $\ denotes saturated ferromagnetic state with
all spins down. For other towers, it is hard to get the explicit form of the
eigenstates, which are dependent of structure of the cluster. In particular,
there is a set of towers with $l=0$, which essentially do not belong to
tower since the corresponding tower length is $1$. However, it does not
affect the statistics of whole energy levels, since the total number of such
energy levels is $N_{\mathrm{tower}}(N/2)=C_{N}^{N/2}-C_{N}^{N/2-1}$,\ which
is a vanishing portion to the whole number of energy levels $2^{N}$, as $N$
tends to infinity.

In this sense, for generic values of $\left\{ J_{ij}\right\} $, the model $%
H_{0}$ is nonthermalizable, even in the case with a set random $\left\{
J_{ij}\right\} $. This can be verified by examining the probability
distribution of the spacings between energy levels (see next section). In
this work we pose the question of whether a perturbation can break the
integrability, and, if it can, is there any towers still remain as quantum
scars.

\section{Transitions of energy level statistics}

\label{Transitions of energy level statistics}

\begin{figure*}[tbph]
\begin{center}
\includegraphics[bb=0 0 610 480,width=0.3\textwidth,clip]{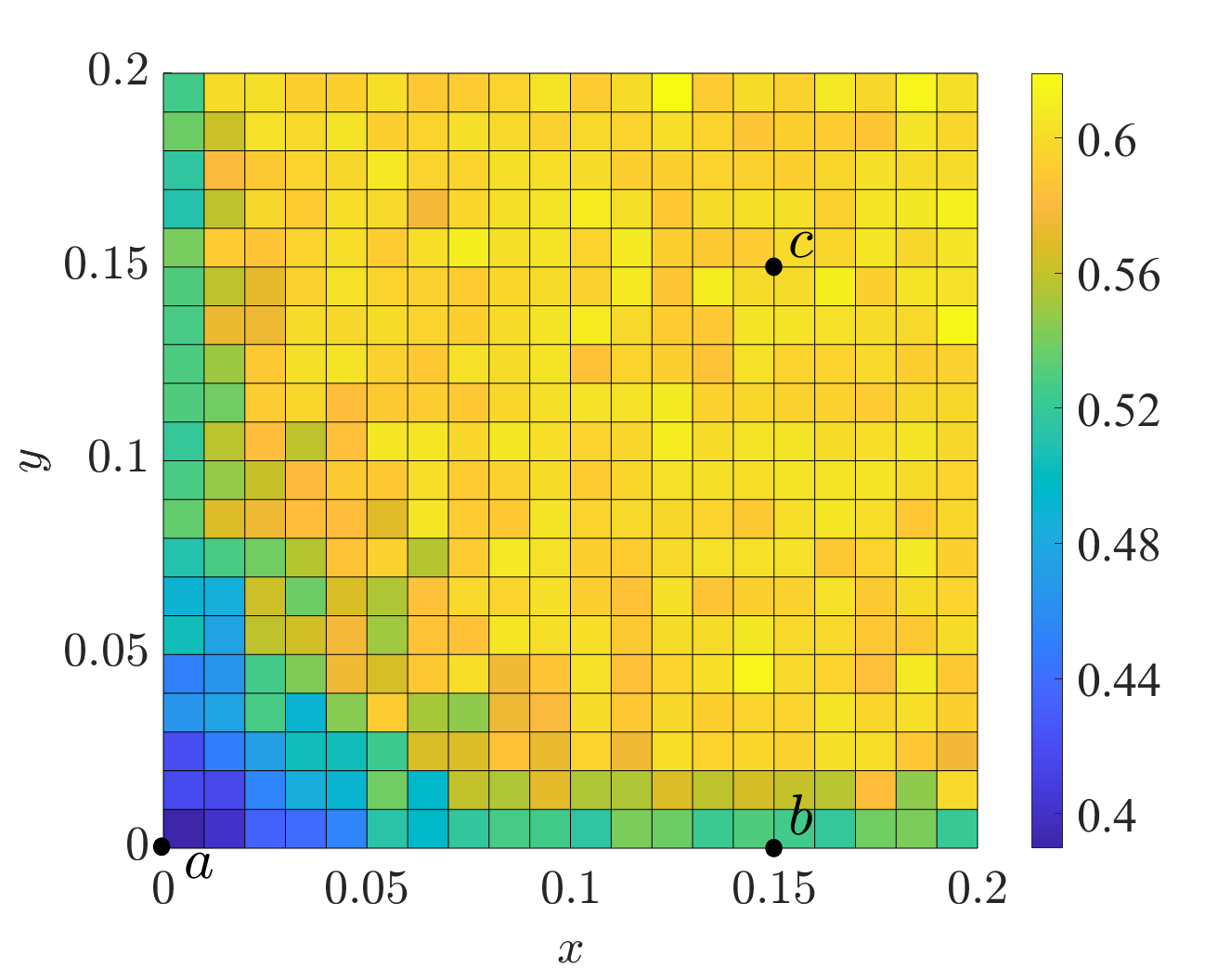} %
\includegraphics[bb=0 0 385 300,width=0.3\textwidth,clip]{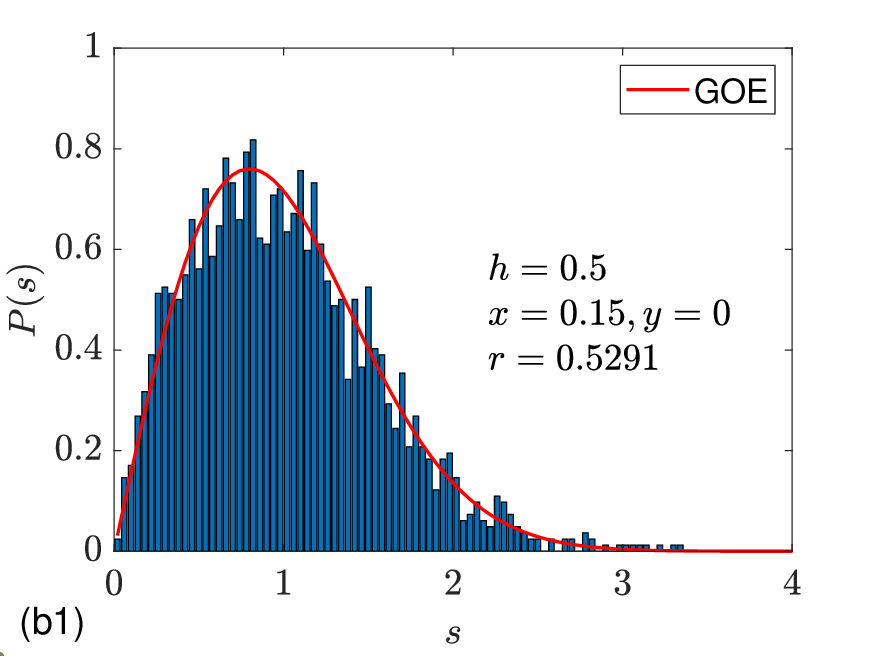} %
\includegraphics[bb=0 0 385 300,width=0.3\textwidth,clip]{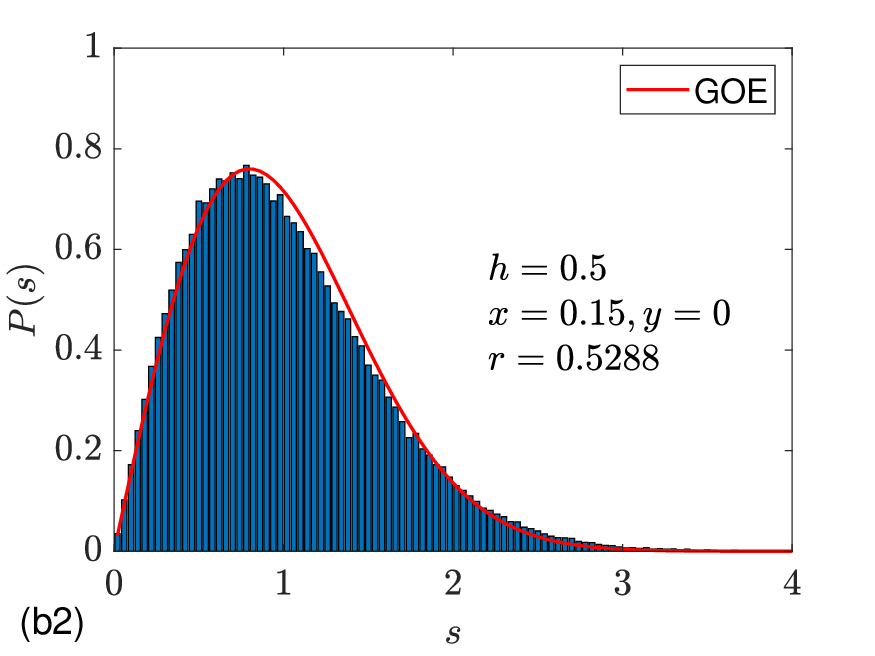} %
\includegraphics[bb=0 0 385 300,width=0.3\textwidth,clip]{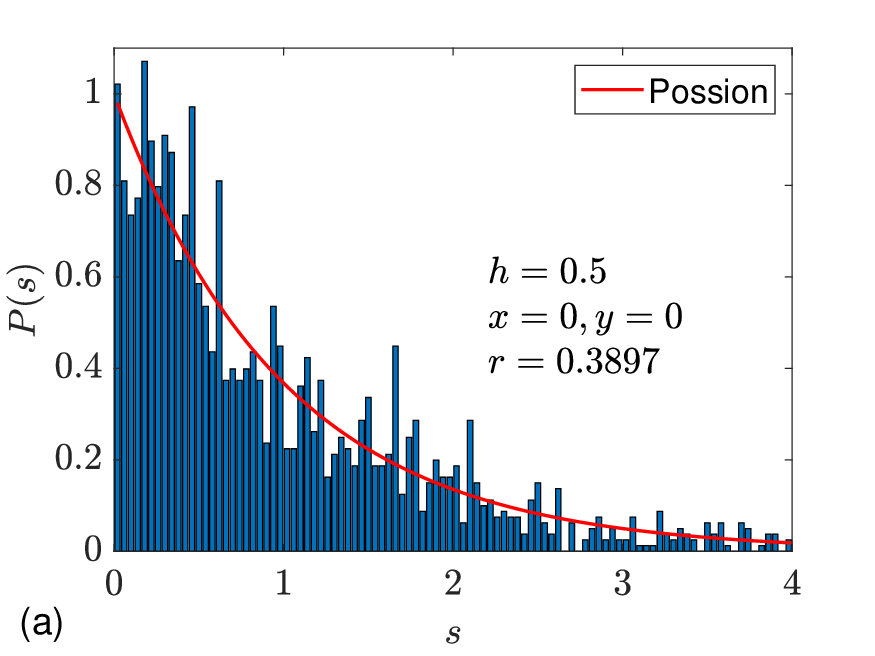} %
\includegraphics[bb=0 0 385 300,width=0.3\textwidth,clip]{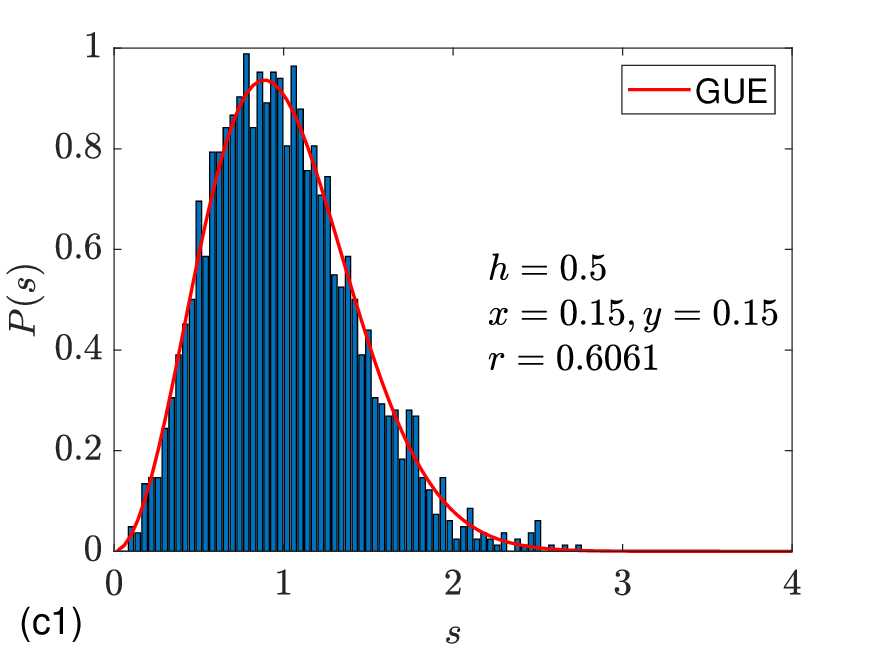} %
\includegraphics[bb=0 0 385 300,width=0.3\textwidth,clip]{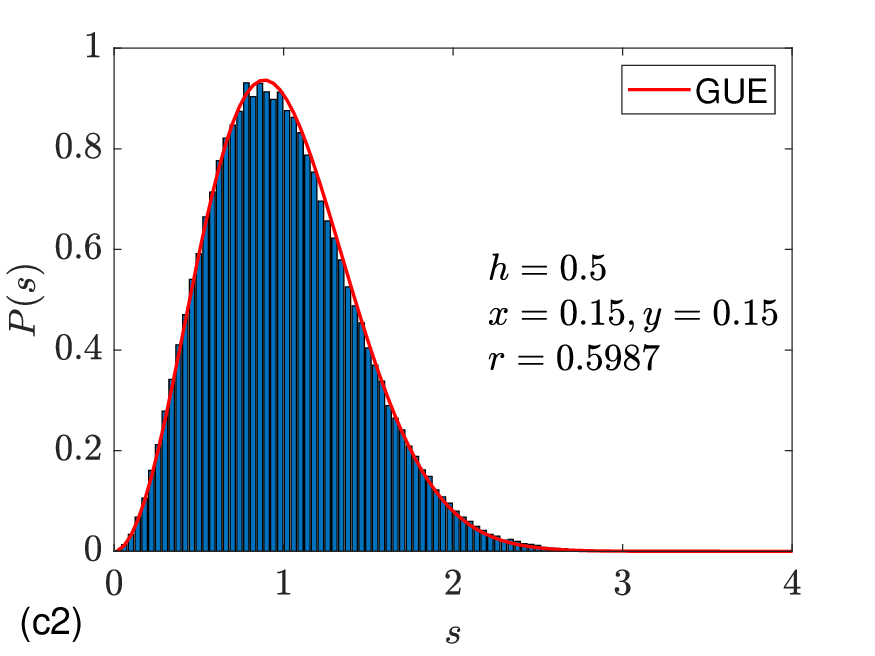}
\end{center}
\caption{Exact diagonalization results on $r$-value and\ statistics of
energy level spacings for the model on $N=12$ chain (\protect\ref{H}) with
representative parameters. (a) Colour contour plot of $r$-value as functions
of ($x,y$) for a given $h_{z}$, obtained from the exact diagonalization of
the system with a single set of random number $\left\{ x_{j}\right\} $ and $%
\left\{ y_{j}\right\} $\ for given ($x,y$). (a, b1, c1) Plots of P(s) for
three typical points in $xy$-plane. The results are obtained from a a single
set of random number. (b2, c2) Same plots of results obtained from the
average over $100$ sets of random number. The Hamiltonian parameters and $r$%
-value used are indicated in the figure. The color lines indicates Poisson,
WD-GOE, and WD-GUE distributions for comparison, the characteristic of
integrable and chaotic systems described by random matrix theory. Excellent
agreement with the typical distributions is shown, especially for the
results from average scheme.}
\label{figure1}
\end{figure*}

\begin{figure*}[tbph]
\begin{center}
\includegraphics[bb=0 0 420 320,width=0.6\textwidth,clip]{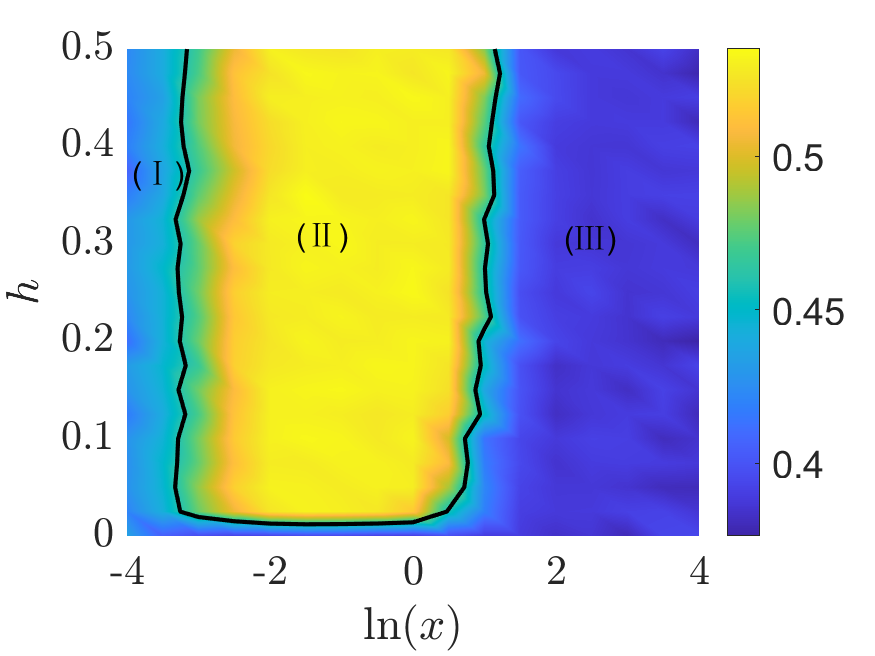}
\end{center}
\caption{Color contour plot of the average level spacing ratio $r$ as
function of $h$ and $x$. There are three regions separated by the black
contour line. The distribution of $r(h,x)$ indicates the phase diagram:
Region (I) is integrable phase, region (II) is non-integrable phase and
region (III) is AL phase. It also indicates that it is always integrable
phase in the vicinity of zero $h$. The results are obtained from the average
over 10 sets of random number.}
\label{figure1.5}
\end{figure*}

According to the above analysis of towers, almost each eigenstate $%
\left\vert \psi _{n}(l,m)\right\rangle $\ of $H_{0}$\ has its own exclusive
set of conserved quantum numbers $(n,l,m)$. Thus in the absence of $H_{%
\mathrm{ran}}$, the system is integrable. In this section, we consider the
case with nonzero $x$ and $y$. When the random transverse fields switch on,
it spoils all the symmetries related the commutation relations in Eqs (\ref%
{SU2}). In the following, we focus on the questions (i) whether the external
field can break the integrability of the system, and (ii) if so, is there
towers can prevent the thermalization.

In spite of the absence of exact solutions, this problem can be investigated
from numerical simulations, e.g., by examining the probability distribution
of the spacings $s$ between energy levels, $P(s)$, which appears to be well
described by random matrix theory for whole levels. First of all, the above
property of towers yields the following conclusions for two extreme cases:
(i) $x=y=0$, but arbitrary $h$; (ii) $x,y\ll \left\vert h\right\vert $,
where the contributions from $H_{\mathrm{ran}}$\ are suppressed
sufficiently. In both cases, all the energy levels in each invariant
subspace with fixed $m$ are just shifted by amount $mh$. Therefore, the
distribution $P(s)$\ in each sector indexed by $m$\ is Poisson distribution.
For the case with nonzero $x$ and $y$, the sectors with different $m$\ are
hybridized, and then one has to count $P(s)$\ for whole levels.\ In Fig. (%
\ref{figure1}), we plot the energy level spacing statistics of the model
with finite $N$. We find that the level spacing distribution is dependent of
three parameters $\left( x,y,h\right) $: (i) $\left( x,y,h\right) $ $=\left(
0,0,0.5\right) $, $P(s)$\ is the Poisson distribution as expected; (ii) $%
\left( x,y,h\right) $ $=\left( 0.15,0,0.5\right) $ or symmetrically $\left(
0,0.15,0.5\right) $, $P(s)$\ is the Wigner-Dyson from the Gaussian
Orthogonal Ensemble (WD-GOE) distribution; (iii) $\left( x,y,h\right) $ $%
=\left( 0.15,0.15,0.5\right) $, $P(s)\ $is the Wigner-Dyson from Gaussian
Unitary Ensemble (WD-GUE) distribution. Such two distributions is typical of
nonintegrable models\cite{Oliviero2021}. The appearence of GOE and GUE
distributions accords with the random matrices theory\cite{Alessio2016},
that different types of variables in a random matrix can lead to different
distribution forms. If all entries in the Hamiltonian are real and satisfy $%
H_{ij}=H_{ji}$, the system exhibits the GOE distribution, while the GUE if
the entities are complex and satisfy $H_{ij}=H_{ji}^{\ast }$.

Another standard numerical test of integrability is to compute the average
level spacing ratio $r$-value\cite{Oganesyan2007}. For a selected set of
energy levels $\left\{ E_{l}\right\} $ (e.g. including the whole levels, or
levels in a certain sector), $r$ is the ratio of adjacent gaps as
\begin{equation}
r_{l}=\frac{\mathrm{min}\{s_{l},s_{l+1}\}}{\mathrm{max}\{s_{l},s_{l+1}\}},
\end{equation}%
and average this ratio over $l$, where $s_{l}$ is level spacings $%
E_{l}-E_{l-1}$. In general, WD-GOE and WD-GUE distributions correspond to $%
r\approx 0.53$ and $r\approx 0.60$ respectively, while Poisson distribution
is $r\approx 0.39$. We introduce a slightly perturbation in numerical
calculations to avoid numerical difficulties arising from degenerate
eigenstates, such as superposition of degenerate eigenstates or the
occurrence of $0/0$ and so on. In Fig. (\ref{figure1.5}), we also plot a
phase in terms of average level spacing ratio $r$ in $h$ vs disorder
strength $x$ plane. As can be seen from the figure, there is a crossover
from integrable to non-integrable phase, and an emergence of Anderson
localization (AL) phase when $x$ is large. There are three regions: (I) and
(III) are integrable phases while (II) is non-integrable phase. The strong
disorder region (III) is Anderson localization (AL) phase\cite{We}. And now,
we concentrate on the non-integrable phase which $x=0.15$. Fig. (\ref%
{figure2}a-c), we plot the $r$-value as functions of ($x,y$) for three types
of lattices, which accord to the plots of $P(s)$.

\begin{figure*}[tbph]
\begin{center}
\includegraphics[bb=0 0 385 300,width=0.3\textwidth,clip]{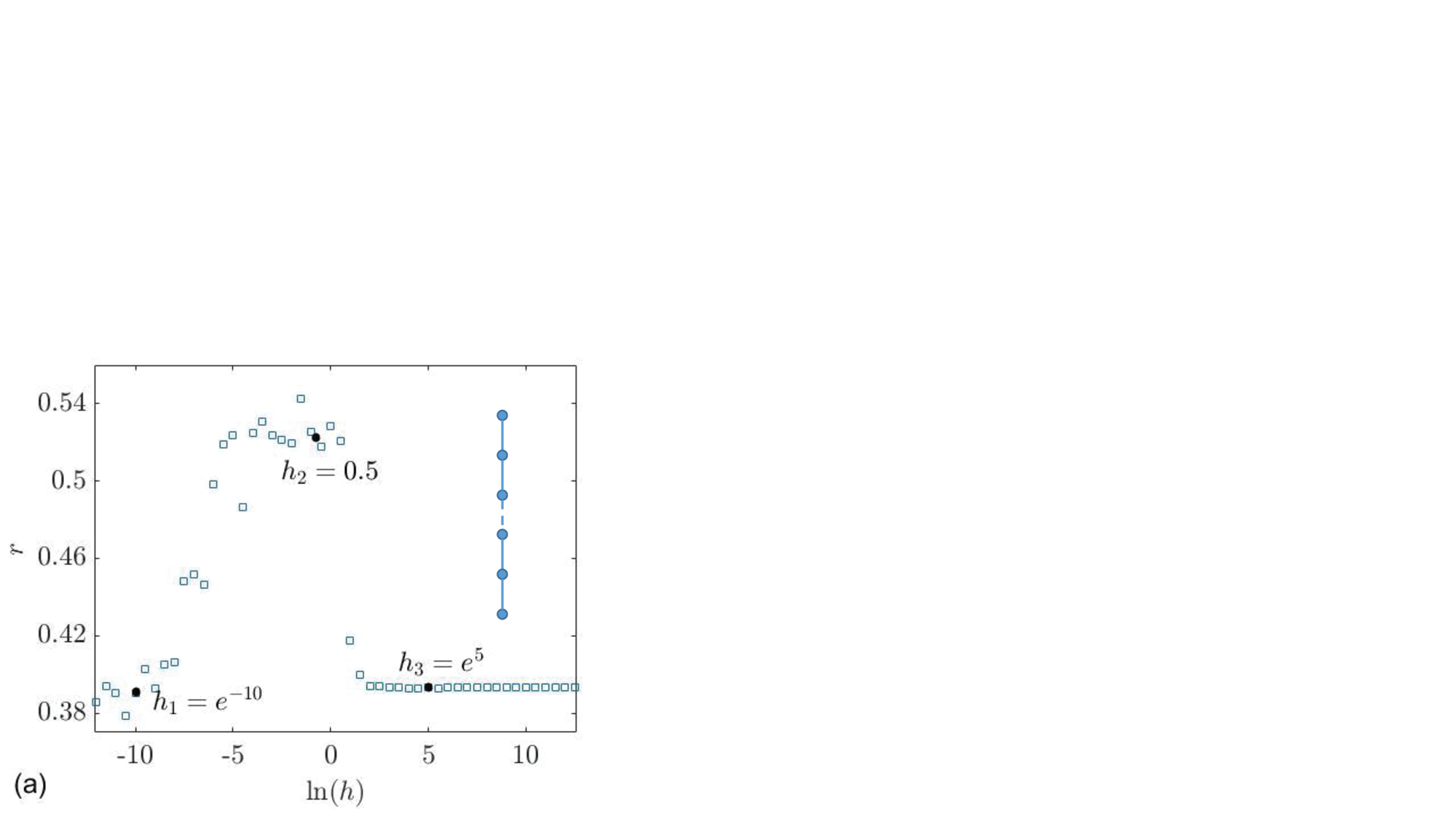} %
\includegraphics[bb=0 0 385 300,width=0.3\textwidth,clip]{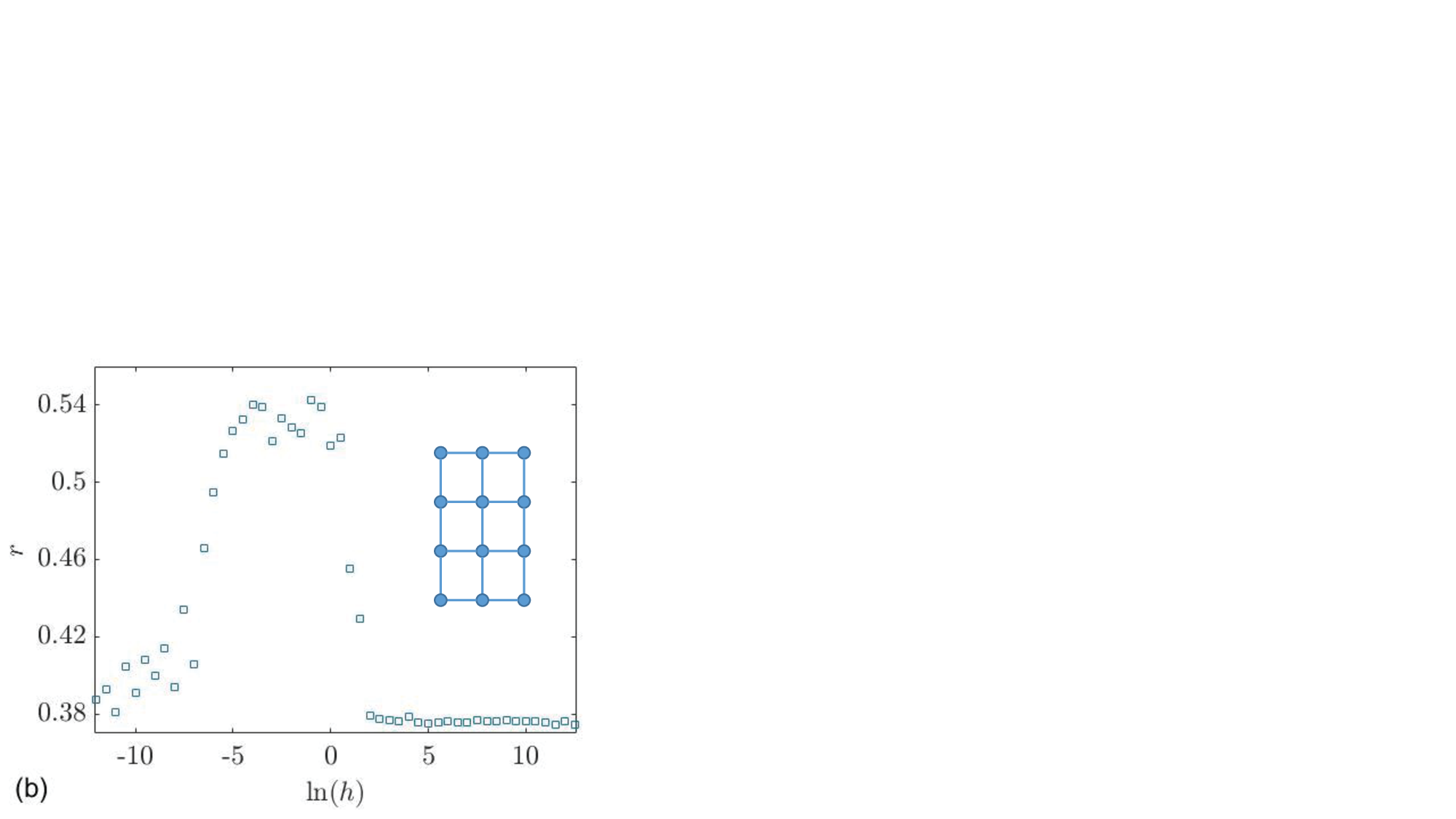} %
\includegraphics[bb=0 0 385 300,width=0.3\textwidth,clip]{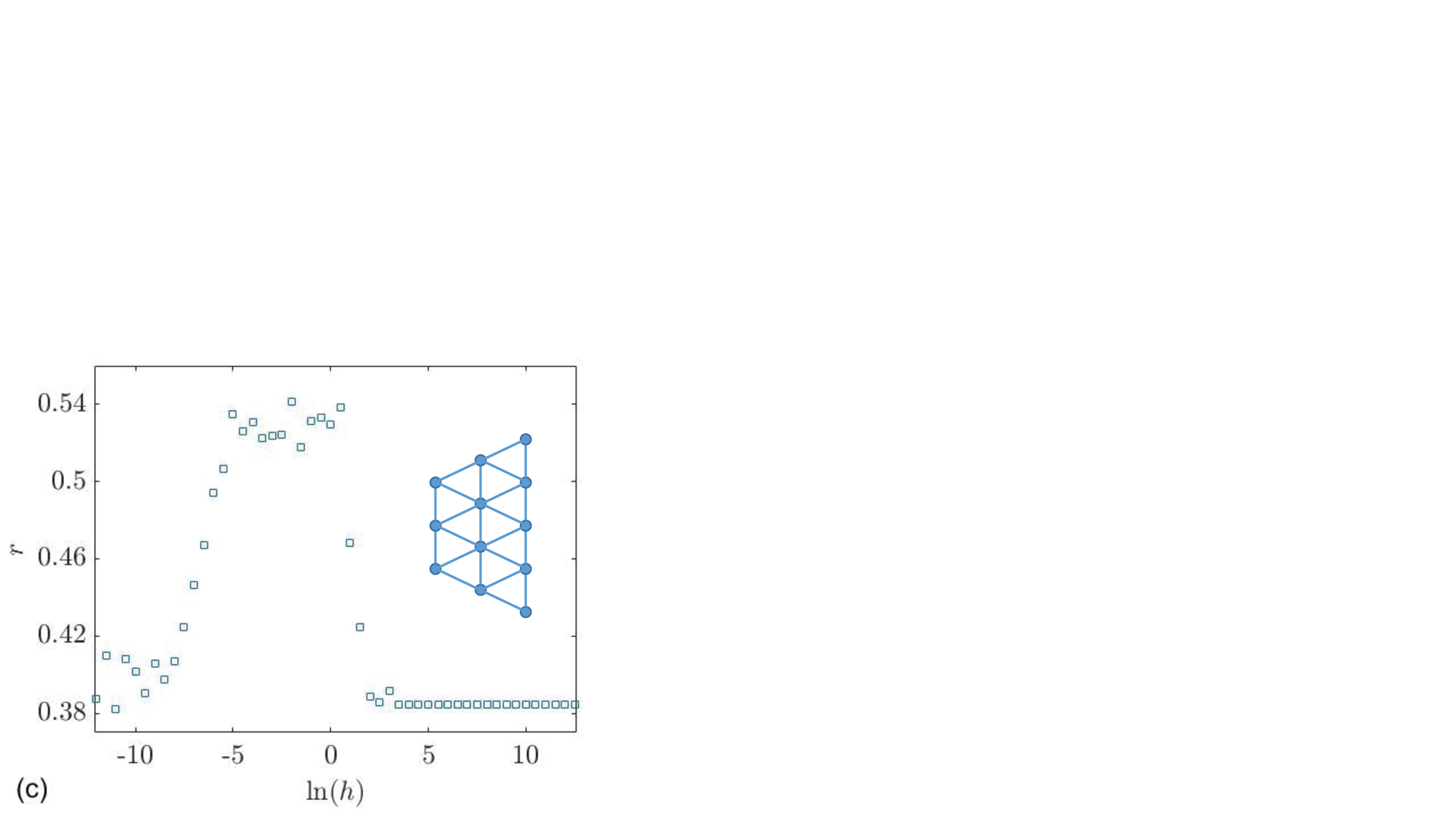} %
\includegraphics[bb=0 0 385 300,width=0.3\textwidth,clip]{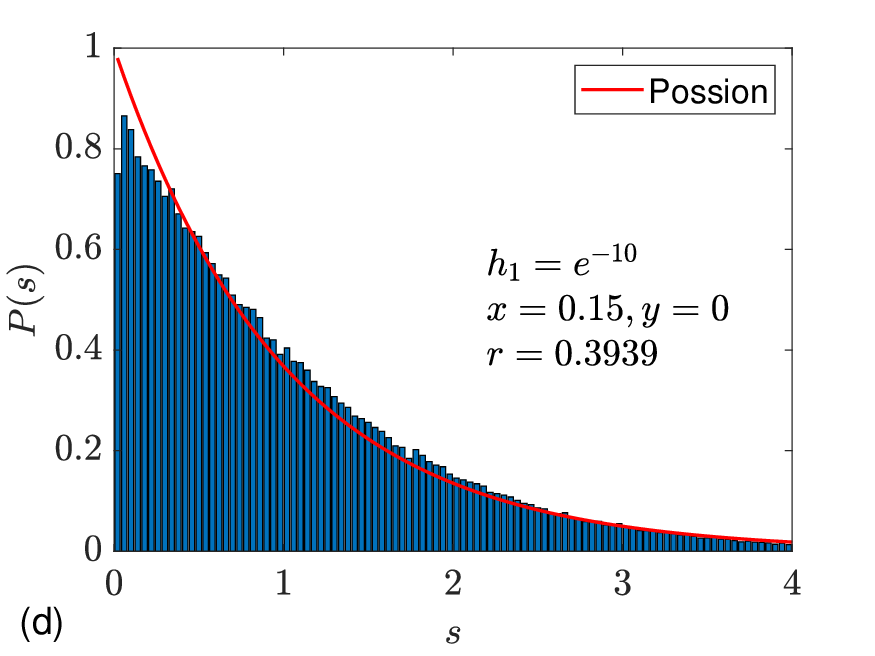} %
\includegraphics[bb=0 0 385 300,width=0.3\textwidth,clip]{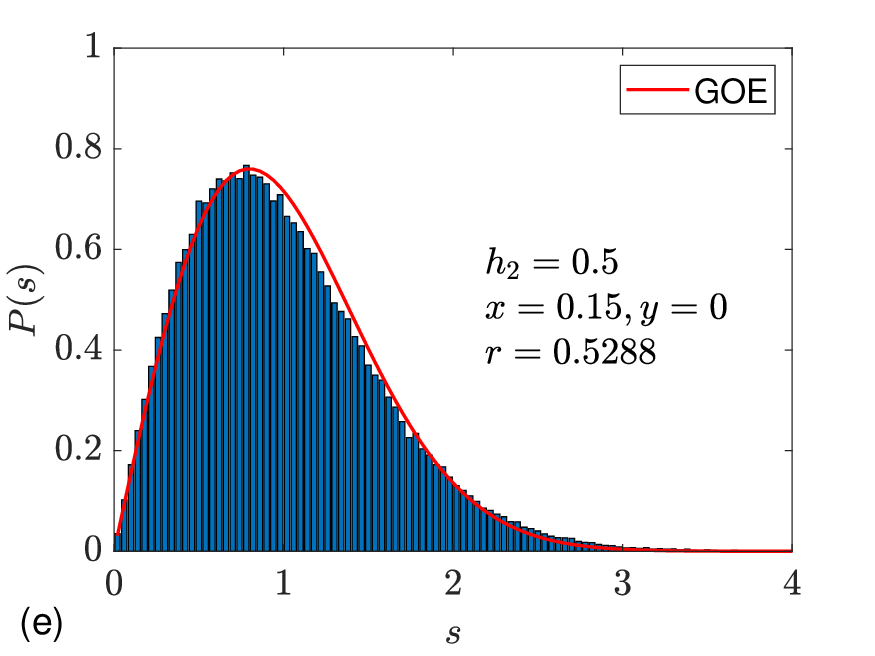} %
\includegraphics[bb=0 0 385 300,width=0.3\textwidth,clip]{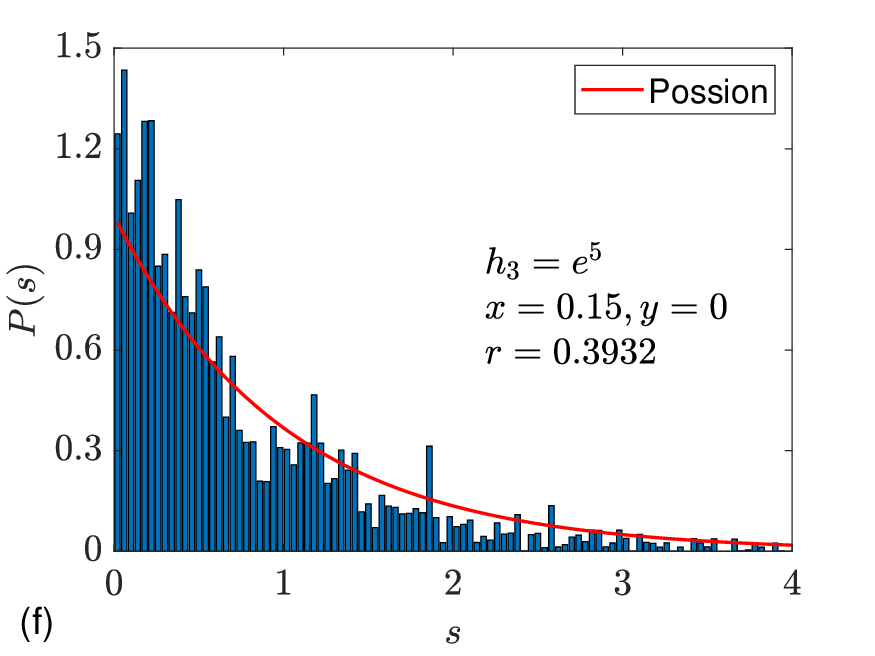}
\end{center}
\caption{The same plots as Fig. (\protect\ref{figure1}) but for three types
of clusters with $N=12$. (a-c) Plots of $r$-value as functions of $h_{z}$
for a given ($x,y$), obtained from the exact diagonalization of the system
with a single set of random number $\left\{ x_{j}\right\} $ and $\left\{
y_{j}\right\} $\ for given ($x,y$). (d-f) Plots of P(s) for three typical
values of $h_{z}$. The results are obtained from the average over $100$ sets
of random number. The Hamiltonian parameters and $r$-value used are
indicated in the figure. The color lines indicates Poisson and WD-GOE
distributions for comparison. Here we only plot the result for the chain,
which has no difference with that from other two clusters.}
\label{figure2}
\end{figure*}

In order to investigate the effect of the uniform field $h$\ on the
transition of energy level statistics of system with nonzero $x$ and $y$, we
plot the $r$-value as function of $h$, and the distributions $P(s)\ $at
representative points in Fig. (\ref{figure2}d-f). We find that the field $h$%
\ takes a subtle role for the integrability of the model. For zero $h$, the
nonzero $x$\ or $y$\ solely cannot induce the non-integrability. On the
other hand, so does the large $h$, which is believed to suppress the random
field.

\section{Quantum scars}

\label{Quantum scars}

\begin{figure*}[tbph]
\begin{center}
\includegraphics[bb=120 238 480 528,width=0.3\textwidth,clip]{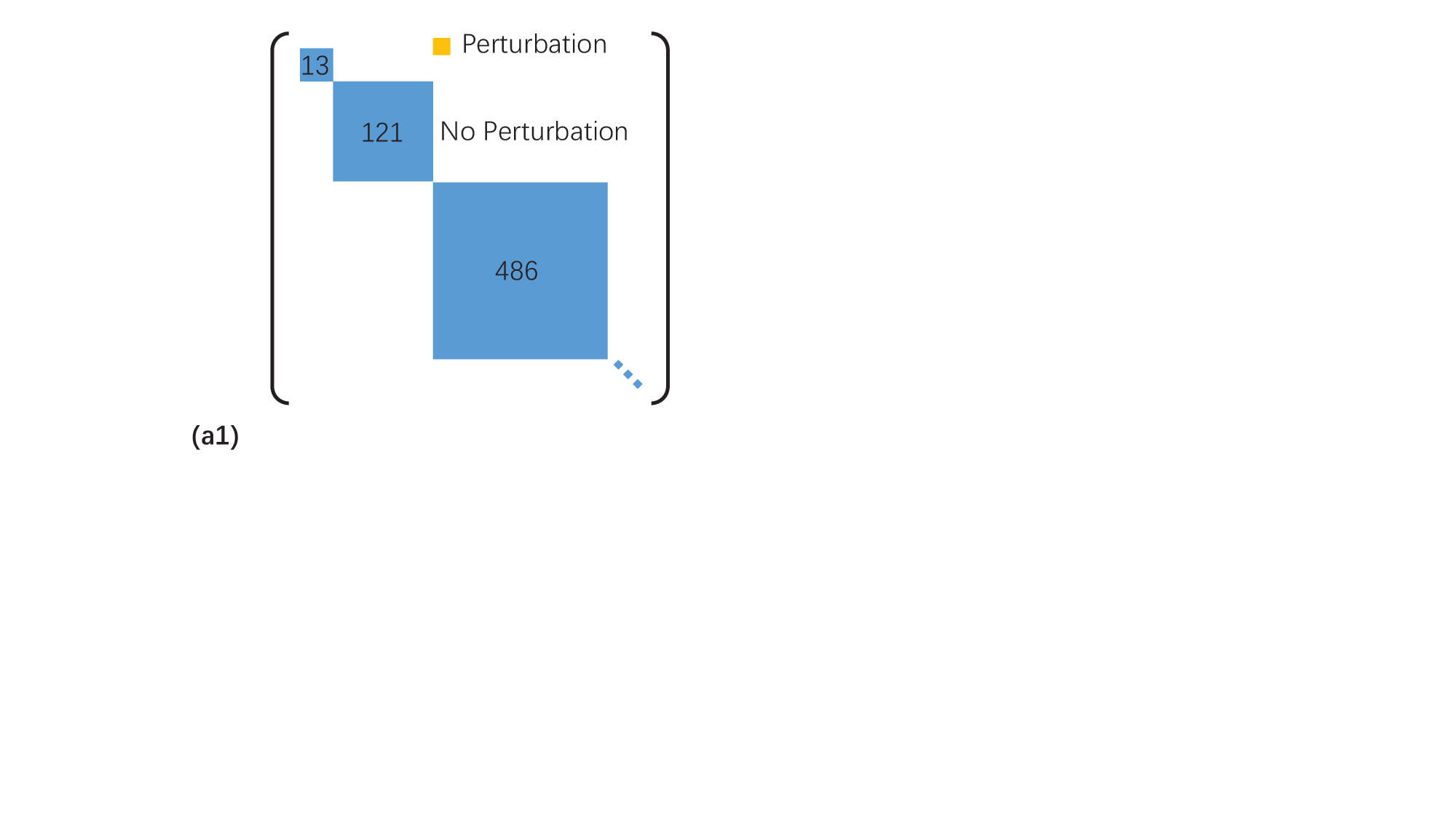} %
\includegraphics[bb=120 12 480 302,width=0.3\textwidth,clip]{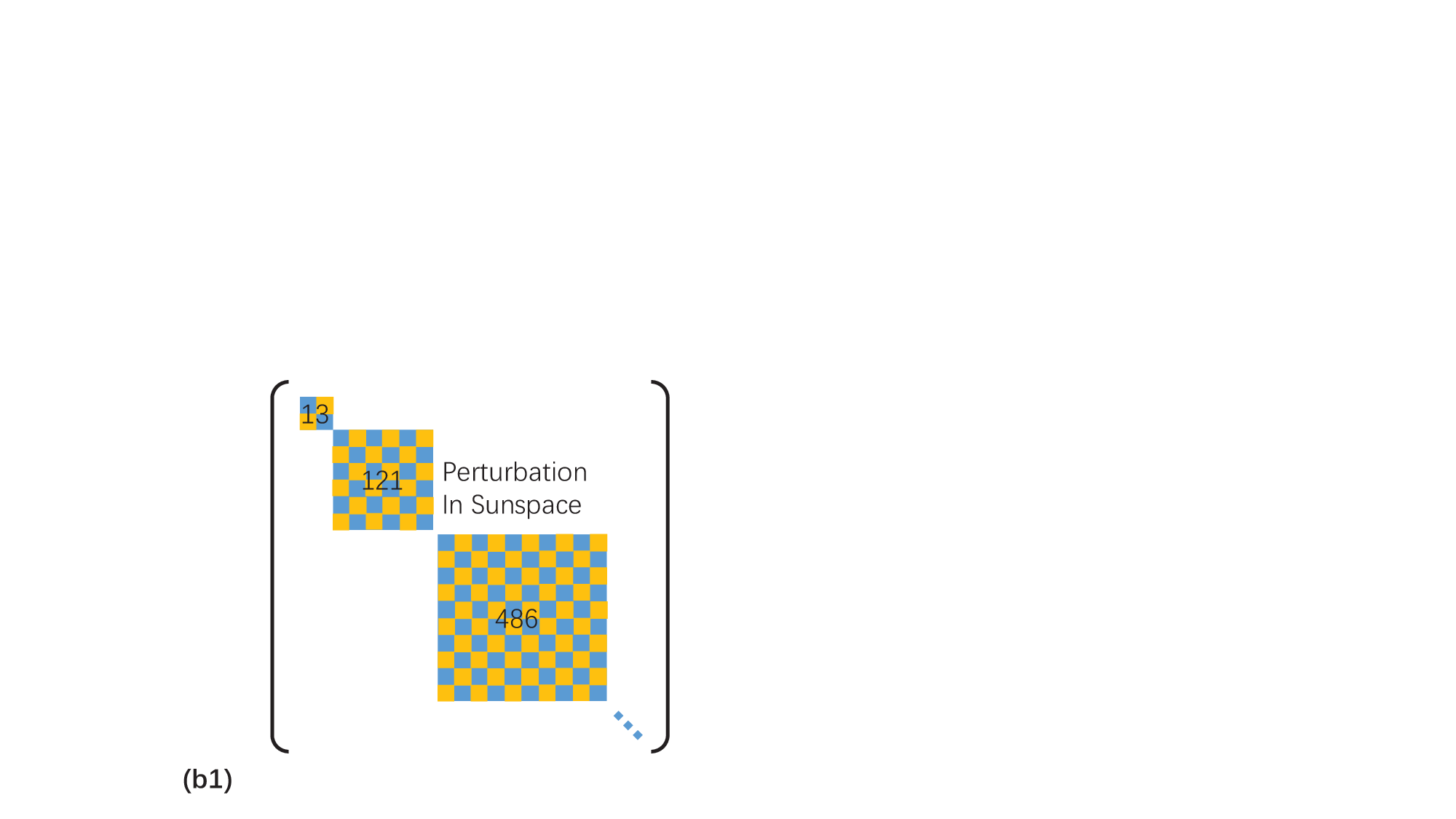} %
\includegraphics[bb=120 238 480 528,width=0.3\textwidth,clip]{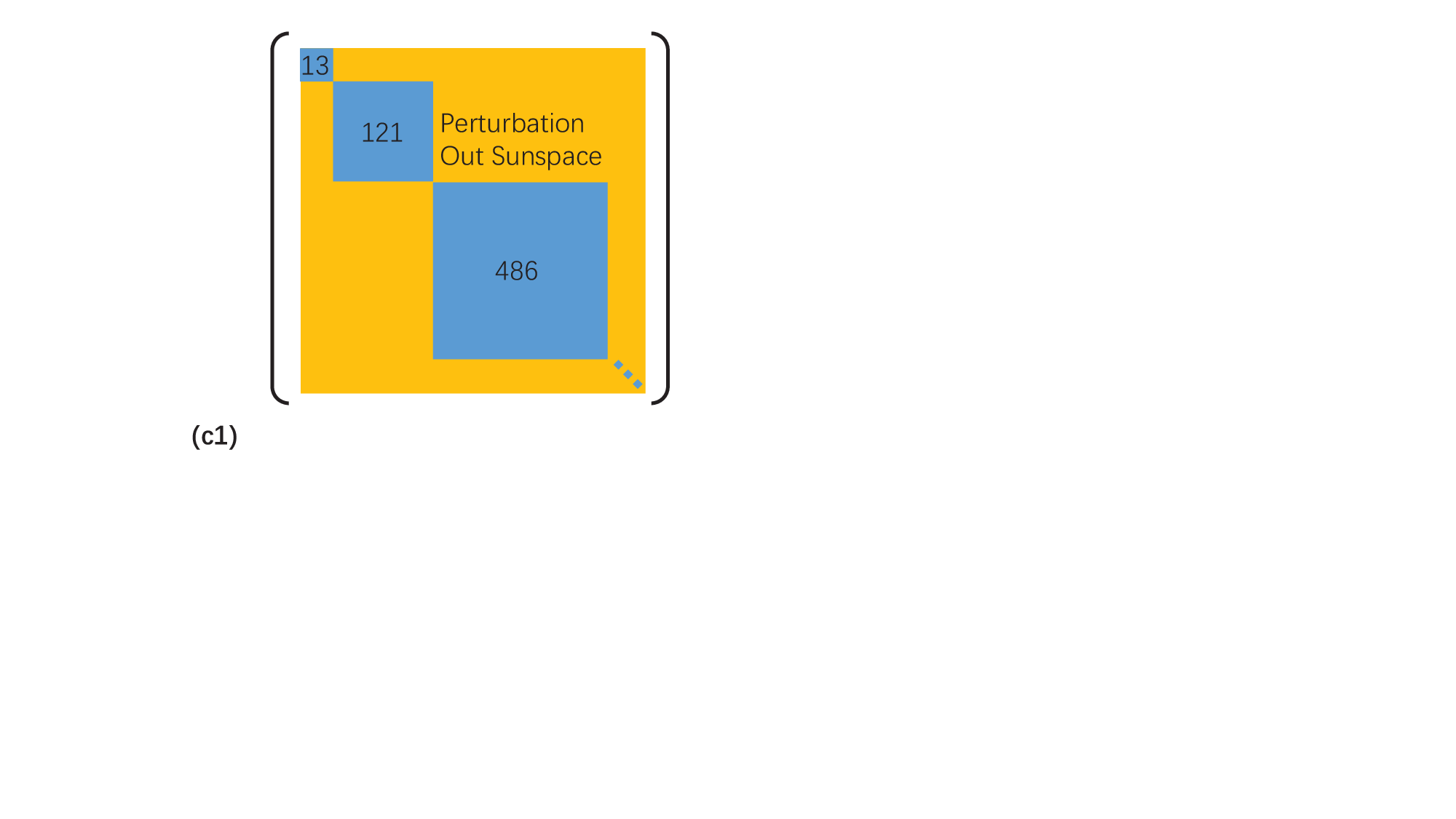} %
\includegraphics[bb=5 0 385 300,width=0.3\textwidth,clip]{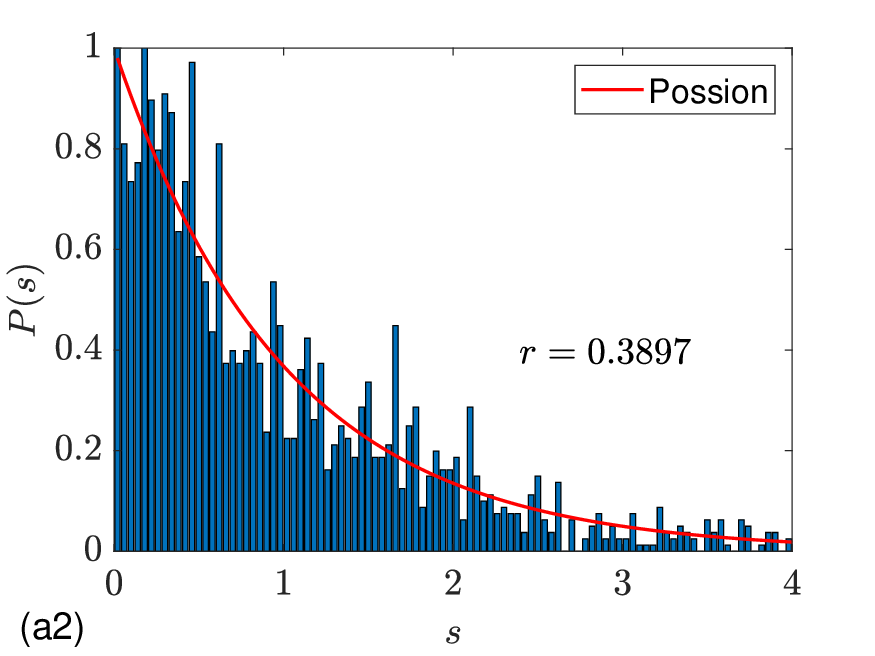} %
\includegraphics[bb=5 0 385 330,width=0.3\textwidth,clip]{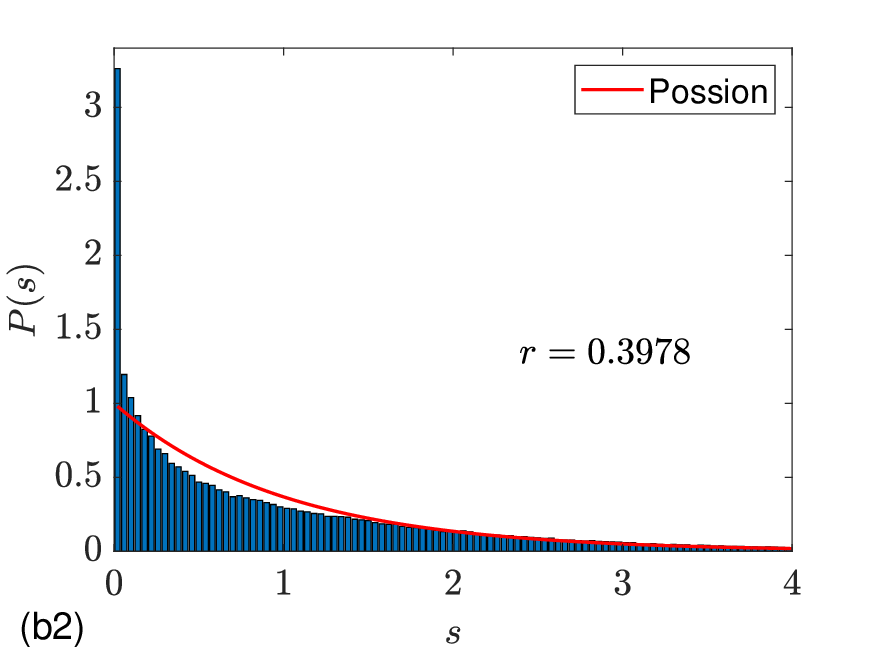} %
\includegraphics[bb=5 0 385 300,width=0.3\textwidth,clip]{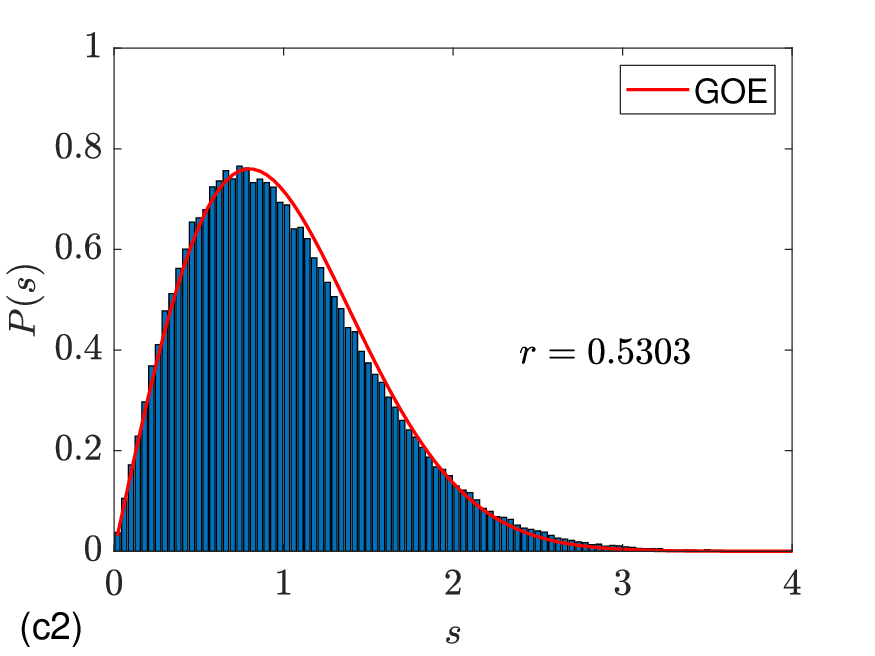}
\end{center}
\caption{Schematic illustration for structures of the matrices, representing
the Hamiltonians with different types of perturbations. (a1) No perturbation
with zero $x$\ and $y$. The blue blocks represent the matrices in different $%
l$ sectors, which consist of towers of eigenstates with different $m$. (b1)
Switching on the perturbations between towers within each $l$ sector. (c1)
Switching on the perturbations between sectors but off perturbations between
towers within each $l$ sector. (a2-c2) The corresponding $P(s)$ obtained
from the exact diagonalization of the matrices (a1-c1). The plots are
obtained from the average over $100$ sets of random number. The parameters
and $r$-value used are indicated in the figure. It indicates that the
perturbations between towers\ with different sectors take the crucial role
for the transition from integrability to non-integrability of the system.}
\label{figure3}
\end{figure*}

In this section, we concentrate on the mechanism of the transition from
integrability to non-integrability induced by external field and identify
the quantum scars. We note that the effects of $H_{\mathrm{ran}}$ on the
states $\left\{ \left\vert \psi _{n}(l,m)\right\rangle \right\} $ are two
classes: (i) hybridizing the levels with the same $l$, by the nonzero
element $\left\langle \psi _{n}(l,m)\right\vert H_{\mathrm{ran}}\left\vert
\psi _{n^{\prime }}(l,m\pm 1)\right\rangle $; (ii) hybridizing the levels
with different $l$, by the nonzero element $\left\langle \psi
_{n}(l,m)\right\vert H_{\mathrm{ran}}\left\vert \psi _{n^{\prime
}}(l^{\prime },m\pm 1)\right\rangle $ with $(l\neq l^{\prime })$. Hence a
natural question arises that which types of elements take the role on the
transition of the level statistics. To answer this question, we consider two
matrices by imposing $\left\langle \psi _{n}(l,m)\right\vert H_{\mathrm{ran}%
}\left\vert \psi _{n^{\prime }}(l,m\pm 1)\right\rangle =0$\ and $%
\left\langle \psi _{n}(l,m)\right\vert H_{\mathrm{ran}}\left\vert \psi
_{n^{\prime }}(l^{\prime },m\pm 1)\right\rangle =0$, respectively. In Fig. (%
\ref{figure3}) we schematically illustrate the structures of the matrices
and plot the corresponding distributions $P(s)$\ in comparison to the exact
one. It evidently shows that\ the hybridization between towers with
different $l$ is determinant for the transition of the level statistics.

\begin{figure}[tbph]
\begin{center}
\includegraphics[bb=5 0 400 300,width=0.4\textwidth,clip]{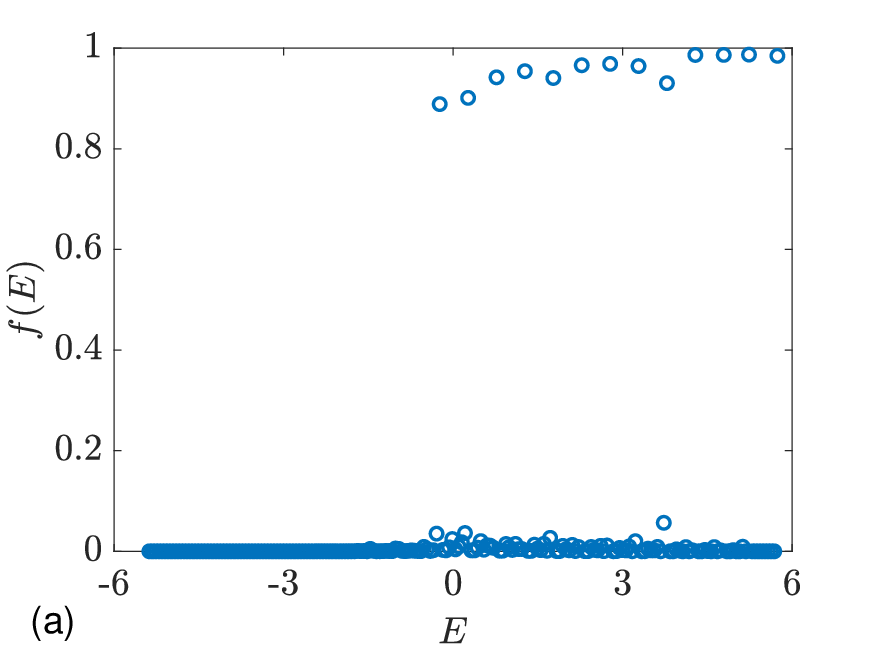} %
\includegraphics[bb=5 0 400 300,width=0.4\textwidth,clip]{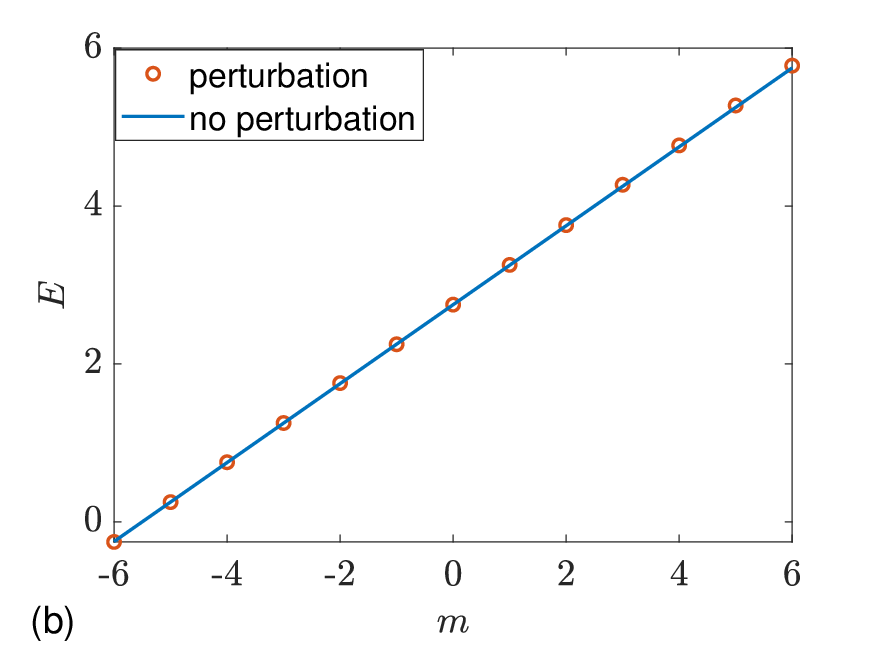}
\end{center}
\caption{Numerical results obtained by exact diagonalization demonstrating
the surviving tower. (a) Plot of $f(E)$\ from (\protect\ref{f(E)}) to reveal
the surviving tower by the peaks. (b) Comparison between the locations of
the peaks with $E_{1}(N/2,m)$. The parameters are $N=12,x=0.15,y=0,$ and $%
\Delta =\left( E_{\max }-E_{\min }\right) /200\approx 0.056.$ This indicates
that the first tower is immune to the perturbation approximately and then
avoids the fast thermalization.}
\label{figure4}
\end{figure}

This result indicates that most of the towers are destroyed by the term $H_{%
\mathrm{ran}}$. Now we consider the question whether there are some towers
surviving from the random perturbation. To this end, we perform numerical
simulation to investigate the effect of $H_{\mathrm{ran}}$\ on the
individual tower. In Fig. (\ref{figure4}) we plot the perturbed levels in
the tower with $l=N/2$. We find that the energy levels are slightly changed,
maintaining the equal spacing very well as quantum scar.

In order to measure the fidelity of the quantum scar of the tower $%
\left\vert \psi _{n}(l,m)\right\rangle $, we introduce the quantity as
function of energy%
\begin{equation}
f(E)=\sum_{E-\Delta /2}^{E+\Delta /2}\sum_{m=-l}^{l}\left\vert \langle \phi
(E^{\prime })\left\vert \psi _{n}(l,m)\right\rangle \right\vert ^{2}
\label{f(E)}
\end{equation}%
for a set of eigenstates $\left\{ \left\vert \phi (E)\right\rangle \right\} $
of $H$, i.e., $H\left\vert \phi (E)\right\rangle =E\left\vert \phi
(E)\right\rangle $. Here we take small $\Delta $\ to select the
quasi-degenerate states near the tower energy levels. Obviously, for perfect
quantum scar, where $\left\{ \left\vert \phi (E)\right\rangle \right\} $\
contains $\left\{ \left\vert \psi _{n}(l,m)\right\rangle \right\} $, we
should have
\begin{equation}
f(E)=\sum_{E-\Delta /2}^{E+\Delta /2}\sum_{m=-l}^{l}\delta \left[ E^{\prime
}-E_{n}(l,m)\right] ,
\end{equation}%
i,e., $f(E)=1$\ when $E=E_{n}(l,m)$. In the presence of nonzero $H_{\mathrm{%
ran}}$, the peaks of $f(E)$\ indicate and measure the efficiency of the
surviving towers for a given perturbed $H$. Numerical simulation is
performed for the tower $\left\vert \psi _{1}(6,m)\right\rangle $\ as an
example. Based on the results of exact diagonalization, $2\times 6+1$\ peaks
of $f(E)$\ is obtained and their positions correspond to the energy levels
of the surviving tower.\ In Fig. (\ref{figure4}), we plot $f(E)$\ and its
peaks to compare the energy ladder for finite system to demonstrate the
surviving tower. We find that the peaks of $f(E)$\ approach to $1$\ for high
energy levels and are more than $0.83$\ for the rest. The energy levels of
the surviving tower have a slight deviation from the exact ladder.\ The
results indicate that the surviving tower is immune to the perturbation
approximately and then avoids the fast thermalization. The behavior of $f(E)$%
\ in this example can be used to explain the dynamics for the specific
initial states in the following section. Here we only present the $f(E)$\
for the chain system for the sake of concise presentation. In fact, similar
numerical results are obtained for other two types of clusters.

\section{Revival of W and GHZ states}

\label{Revival of W and GHZ states}

\begin{figure*}[tbph]
\begin{center}
\includegraphics[bb=0 0 400 430,width=0.3\textwidth,clip]{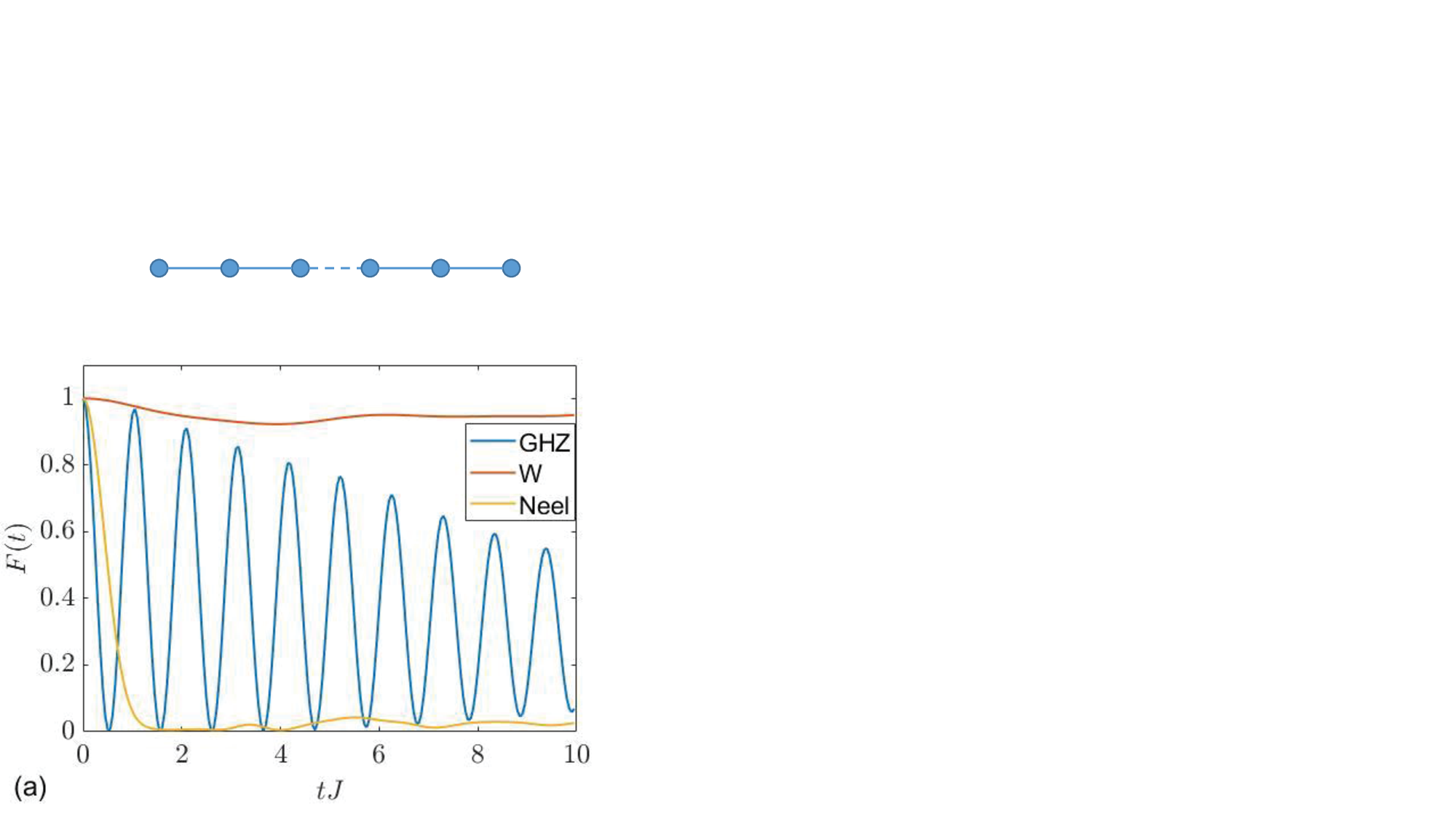} %
\includegraphics[bb=0 0 400 430,width=0.3\textwidth,clip]{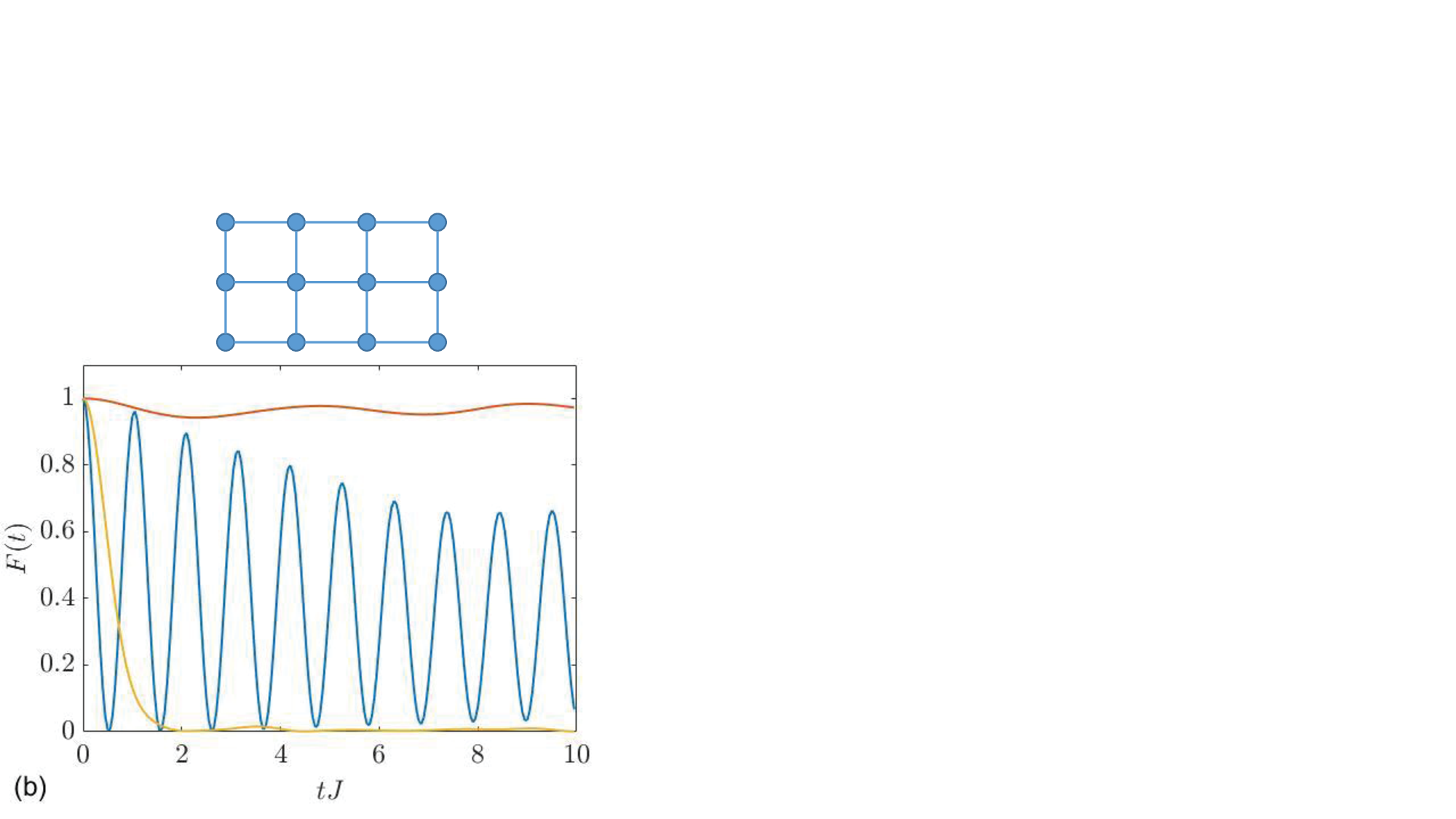} %
\includegraphics[bb=0 0 400 430,width=0.3\textwidth,clip]{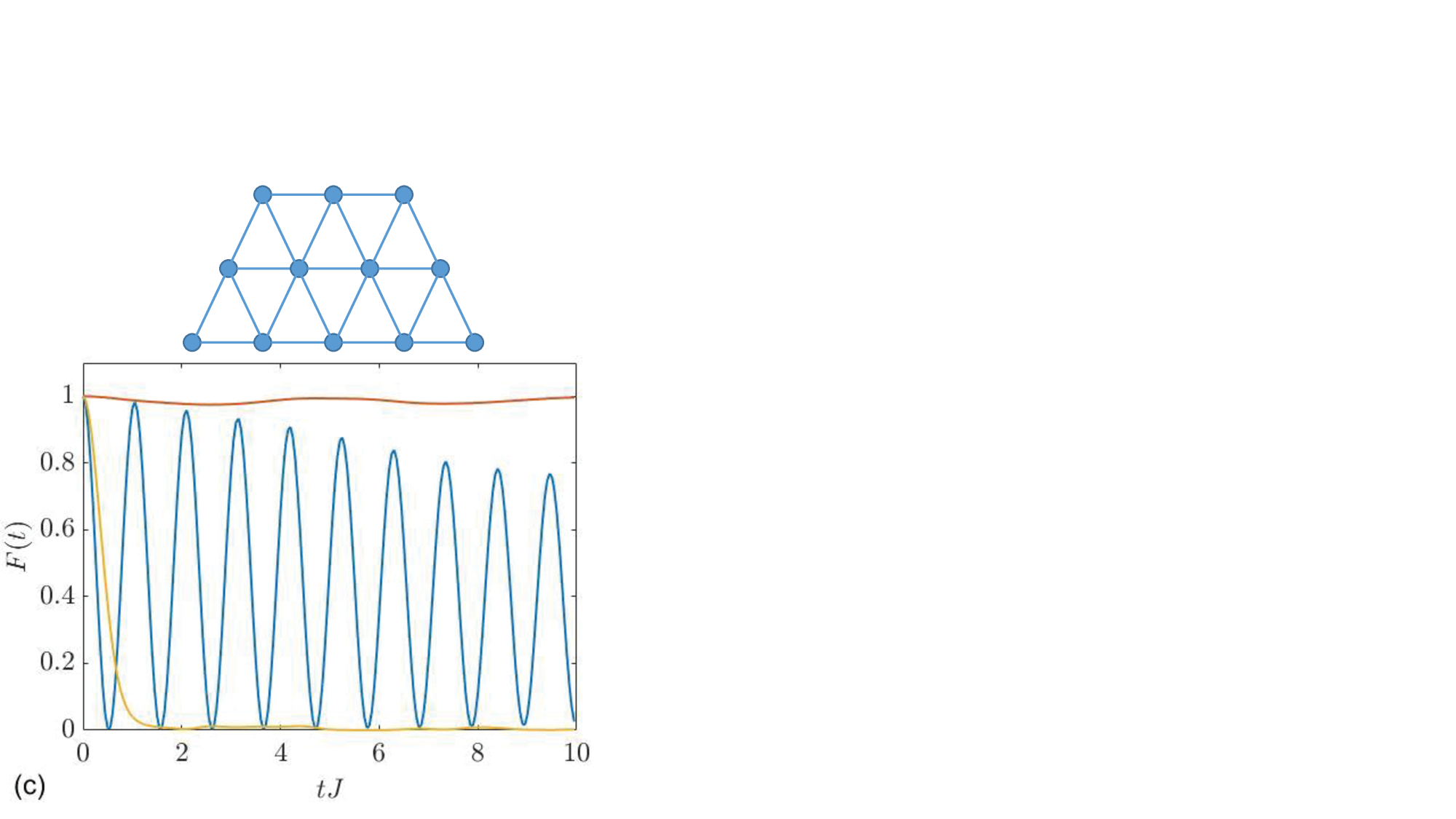}
\end{center}
\caption{Plots of the fidelity from (\protect\ref{Ft}) for three types of
Heisenberg clusters. The initial states are W state $\left\vert \mathrm{W}%
\right\rangle =\left\vert \protect\psi _{1}(6,5)\right\rangle $, GHZ state $%
\left\vert \mathrm{GHZ}\right\rangle $\ $=\frac{1}{\protect\sqrt{2}}\left(
\left\vert \protect\psi _{1}(6,6)\right\rangle +\left\vert \protect\psi %
_{1}(6,-6)\right\rangle \right) $, and Neel state $\left\vert \mathrm{Neel}%
\right\rangle =\left\vert \uparrow \downarrow \uparrow \downarrow \uparrow
\downarrow \uparrow \downarrow \uparrow \downarrow \uparrow \downarrow
\right\rangle $. The parameters are $N=12$, $x=0.15$, $y=0$, and $h=0.5$. We
find that the behaviors of the fidelity are almost independent of the
geometry of the clusters. Analyses and discussions are given in the text.}
\label{figure5}
\end{figure*}

So far we have shown that the external field can\ induce the thermalization
for an isotropic clusters. In addition, such systems host special nonthermal
eigenstates that should support periodic revival. Inspired by recent
experiments with Rydberg atoms, where nonthermal periodic revival dynamics
has been observed for initial Neel state\cite{Bernien2017,Bluvstein2021}, we
will examine the dynamics for three states. They are W state $\left\vert
\mathrm{W}\right\rangle =\frac{1}{\sqrt{N}}Q\left\vert \Uparrow
\right\rangle $, GHZ state $\left\vert \mathrm{GHZ}\right\rangle =\frac{1}{%
\sqrt{2}}\left( \left\vert \Uparrow \right\rangle +\left\vert \Downarrow
\right\rangle \right) $\ and Neel state, which can be expressed in the form%
\begin{equation}
\left\vert \mathrm{W}\right\rangle =\left\vert \psi
_{1}(N/2,N/2-1)\right\rangle
\end{equation}%
and%
\begin{equation}
\left\vert \mathrm{GHZ}\right\rangle =\frac{1}{\sqrt{2}}\left( \left\vert
\psi _{1}(N/2,N/2)\right\rangle +\left\vert \psi _{1}(N/2,-N/2)\right\rangle
\right) ,
\end{equation}%
and%
\begin{equation}
\left\vert \mathrm{Neel}\right\rangle =\left\vert \uparrow \downarrow
\uparrow \downarrow ...\uparrow \downarrow \right\rangle ,
\end{equation}%
respectively. Here states $\left\vert \mathrm{W}\right\rangle $ and $%
\left\vert \mathrm{GHZ}\right\rangle $ are two different typical
multipartite entangled states, which are usually referred to as maximal
entanglement\cite{Migda?2013}. The two states are included in the tower with
$l=N/2$, while the Neel state $\left\vert \mathrm{Neel}\right\rangle $\ is
not. In practice for quantum computing, an initial state made of many-body
scar states repeatedly returns to itself in time evolution, preventing the
loss of quantum information through thermalization. To demonstrate this
point, we compute the time evolution for the initial states $\left\vert \psi
(0)\right\rangle =\left\vert \mathrm{GHZ}\right\rangle $, $\left\vert
\mathrm{W}\right\rangle $ and $\left\vert \mathrm{Neel}\right\rangle $,
under the Heisenberg clusters with finite $N$. We introduce the fidelity%
\begin{equation}
F(t)=\left\vert \left\langle \psi (0)\right\vert e^{-iHt}\left\vert \psi
(0)\right\rangle \right\vert ^{2},  \label{Ft}
\end{equation}%
to measure the feature of revival. As can be seen in Fig. (\ref{figure5}),
the quantity $F(t)$\ shows the periodic revivals for W and GHZ states, but
relatively fast decay for initial Neel state. From the profile of $f(E)$\ in
Fig. (\ref{figure4}a), the behaviors of $F(t)$\ in Fig. (\ref{figure5})\ for
three initial states can be explained as following. (i) The $\left\vert
\mathrm{W}\right\rangle $ state is almost the eigenstate of $H$\ since the
second peak of $f(E)$\ is very close to $1$. It results in $F(t)\approx 1$\
for the time evolution as expected. (ii) The $\left\vert \mathrm{GHZ}%
\right\rangle $\ state is related first and the last peaks of $f(E)$. We
note that the last peak of $f(E)\approx 0.83$. This deviation from $1$\
definitely leads to a light dumping oscillation. (iii) As for the Neel state
$\left\vert \mathrm{Neel}\right\rangle $, it is a component of $\left\vert
\psi _{1}(6,0)\right\rangle $ with\ a very small amplitude
\begin{equation}
\langle \psi _{1}(6,0)\left\vert \mathrm{Neel}\right\rangle =\sqrt{\frac{1}{%
C_{12}^{6}}}\approx 0.033.
\end{equation}%
Although it is related to the middle peak of $f(E)$, it is almost out of the
scar. Then the $F(t)$\ of Neel state decays fast. The above analysis is
based on the result of $f(E)$\ for the chain system. However, the same
conclusion can be obtained for other types of clusters. As expected,
numerical simulations show that the behavior of $F(t)$\ is almost
independent of the geometry of the clusters. In addition, we would like to
point out that, for another W state $\left\vert \mathrm{W}^{\prime
}\right\rangle =\frac{1}{\sqrt{N}}Q^{\dag }\left\vert \Downarrow
\right\rangle $, the fidelity should be not so perfect as that of $%
\left\vert \mathrm{W}^{\prime }\right\rangle $\ due to the shrinkage\ of the
last second peak of $f(E)$. We would like to point out that the profile of $%
f(E)$\ is sensitive to a very small $\Delta $, and the results presented in
Fig. (\ref{figure4}) and (\ref{figure5})\ are selected from the cases with a
finite number of sets of random numbers. It is unavoidable one may get a
very different result by accident.\ We emphasize again that the conclusions
obtained here hold only to small size systems, i.e., it is not sure whether
or not the random field we applied can result in localization for large $N$\
system, preventing the thermalization.

\begin{figure*}[tbph]
\begin{center}
\includegraphics[bb=50 200 470 520,width=0.6\textwidth,clip]{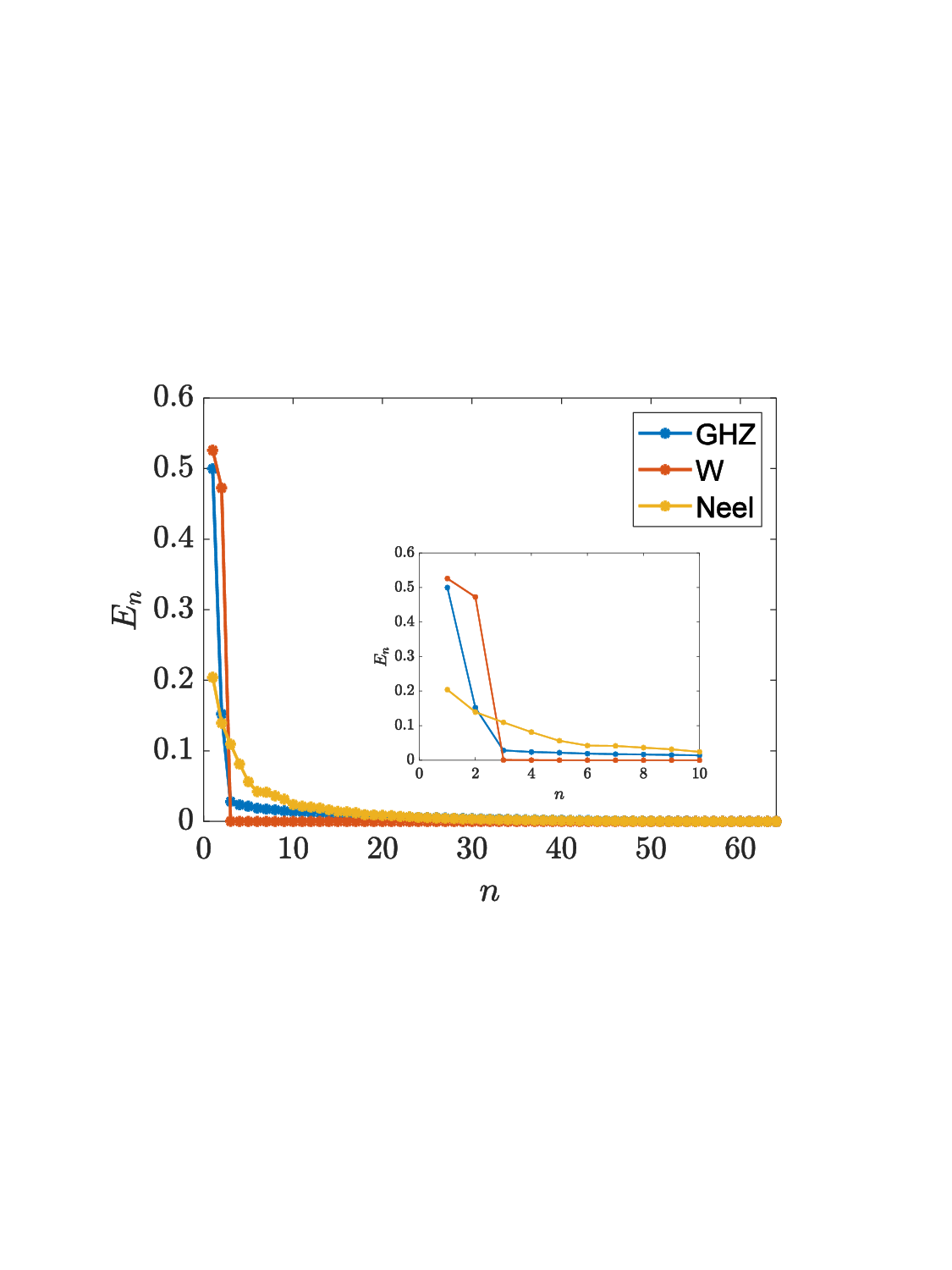}
\end{center}
\caption{Entanglement spectrum (ES) of GHZ, W and Neel states
under long-time evolution ($t=500$) in the regions of non-integrable ($h=0.5$
and $x=0.15 $). In the inset, the large vision of front part of spectrum.
The gaps for GHZ and W states are evident.}
\label{figure6}
\end{figure*}

In addition, we also employ the entanglement spectrum (ES) to support the
existence of quantum scars. The ES has been used to study quantum scars and
distinguish MBL systems from ergodic systems recently\cite%
{Mondal20221,Yang2015,Serbyn2016}. It represents the eigenvalues $\left\{
E_{n}\right\} $ of the reduced density matrix,
\begin{equation}
\rho _{s}=\sum\limits_{n}E_{n}\left\vert n\right\rangle _{s}\left\langle
n\right\vert _{s},
\end{equation}%
which is obtained from the Schmidt decomposition of $\left\vert \psi
\right\rangle =\sum\limits_{n}\sqrt{E_{n}}\left\vert n\right\rangle
_{s}\otimes \left\vert n\right\rangle _{\bar{s}}$. For quantum scars, the ES
will exhibit a large gap. Therefore, we plotted the ES for GHZ, W, and Neel
states in Fig. (\ref{figure6}a). Here we divide the entire system into two
equally sized parts to compute the reduced density matrix. And we found that
the ES of GHZ and W states have a significant gap for sufficiently long time
($t=500$), while the Neel state does not. This is also evidence that they
are quantum scars. On the other hand, recent studies have found
that fidelity out-of-time-order correlator ({FOTOC)} exhibits a non-stable
dynamical behavior with respect to quantum scars\cite{Mondal20222}. However
in our case, the FOTOC, which is defined as $f_{is}=\sum\limits_{\alpha
=x,y,z}\left( \left\langle s_{i\alpha }^{2}\right\rangle -\left\langle
s_{i\alpha }\right\rangle ^{2}\right) /s^{2}$, for the initial W and GHZ
states do not exhibit oscillation behavior. It may be due to the fact that
the W and GHZ states are the quasi-eigenstate and maximally entangled state,
respectively.

\section{Conclusions}

\label{conclusions} We have demonstrated that the external fields, can
induce dramatic transition from integrability to non-integrability for
different clusters of isotropic Heisenberg model. Numerical results show
that the cooperation between the uniform field one direction, and the random
field in other two directions, takes the crucial role for such a transition.
While generic initial states are expected to thermalize, we show that there
is a tower of eigenstates leads to weak ergodicity breaking in the form of
equal level spacing approximately. Specifically, this quantum many-body scar
covers two important states, W and GHZ states. The obtained results indicate
that the random field with moderate strength can induce non-integrability
for finite size clusters. This finding reveals the possibility of quantum
information processing that is immune to the thermalization in finite size
quantum spin clusters at nonzero temperature.

\ack
We acknowledge the support of the National Natural Science Foundation of
China (Grants No. 11705127 and No. 11874225).

\end{document}